\documentclass[12pt,showpacs,prd,preprintnumbers,amsmath,amssymb,floatfix,nofootinbib]{revtex4}
\usepackage{graphicx,dcolumn}

\newcolumntype{d}{D{.}{.}{-1}}
\newcolumntype{e}[1]{D{.}{.}{#1}}
\newcolumntype{f}[1]{D{.}{\mathbf{.}}{#1}}
\newcolumntype{s}[1]{D{/}{\;/\;}{#1}}
\newcommand{\talignc}[1]{\multicolumn{1}{c}{#1}}

\begin{document}
\def\cere{Cherenkov }
\def\numu{${\nu}_{\mu}$}
\def\nue{${\nu}_{e}$}
\def\nutau{${\nu}_{\tau}$}
\def\nuall{${\nu}_{e, \mu, \tau}$}
\def\nuetau{${\nu}_{e, \tau}$}

\title{Five years of searches for point sources of astrophysical neutrinos with the AMANDA-II neutrino telescope}

\author{
A.~Achterberg$^{31}$,
M.~Ackermann$^{33}$\footnote{Corresponding authors: markus.ackermann@desy.de (M. Ackermann) and elisa.bernardini@desy.de (E. Bernardini)},
J.~Adams$^{11}$,
J.~Ahrens$^{21}$,
K.~Andeen$^{20}$,
D.~W.~Atlee$^{29}$,
J.~N.~Bahcall$^{25}$\footnote{Deceased},
X.~Bai$^{23}$
% ...
}

\vspace*{0.2cm}

% ...
\affiliation{$^{11}$Dept.~of Physics and Astronomy, University of Canterbury, Private Bag 4800, Christchurch, New Zealand}
% ...
\affiliation{$^{20}$Dept.~of Physics, University of Wisconsin, Madison, WI 53706, USA}
\affiliation{$^{21}$Institute of Physics, University of Mainz, Staudinger Weg 7, D-55099 Mainz, Germany}
% ...
\affiliation{$^{23}$Bartol Research Institute, University of Delaware, Newark, DE 19716, USA}
% ...
\affiliation{$^{25}$Institute for Advanced Study, Princeton, NJ 08540, USA}
% ...
\affiliation{$^{29}$Dept.~of Physics, Pennsylvania State University, University Park, PA 16802, USA}
% ...
\affiliation{$^{31}$Dept.~of Physics and Astronomy, Utrecht University/SRON, NL-3584 CC Utrecht, The Netherlands}
% ...
\affiliation{$^{33}$DESY, D-15735 Zeuthen, Germany}

\date{\today}
\begin{abstract}
We report the results of a five-year survey of the
northern sky to search for point sources of high energy neutrinos.
The search was performed on the data collected with the AMANDA-II neutrino telescope
in the years 2000 to 2004, with a live-time of 1001 days. 
The sample of selected events consists of 4282 upward going muon tracks with 
high reconstruction quality and an energy larger than about 100 GeV.
We found no indication of point sources of neutrinos and set 90\%
confidence level flux upper limits for an all-sky search and also for
a catalog of 32 selected sources.
For the all-sky search, our average (over declination and right
ascension) experimentally observed upper limit
$\mathrm{\Phi^{0}=\big(\frac{E}{1\,TeV}\big)^\gamma\cdot \frac{d\Phi}{dE}}$ 
to a point source flux of muon and tau
neutrino (detected as muons arising from taus) is
$\mathrm{\Phi_{\nu_\mu+\bar{\nu}_\mu}^\mathrm{0}+\Phi_{\nu_\tau+\bar{\nu}_\tau}^\mathrm{0}}
= 11.1 \cdot 10^{-11} \,\mathrm{TeV}^{-1}\,\mathrm{cm}^{-2}\,\mathrm{s}^{-1}$,
in the energy range between 1.6 TeV and 2.5 PeV for a flavor ratio
$\mathrm{\Phi_{\nu_\mu+\bar{\nu}_\mu}^\mathrm{0}/\Phi_{\nu_\tau+\bar{\nu}_\tau}^\mathrm{0}}= 1$
and assuming a spectral index $\gamma$=2. It should be noticed that this is the first 
time we set upper limits to the flux of muon and tau neutrinos.
In previous papers we provided muon neutrino upper limits only neglecting the sensitivity to 
a signal from tau neutrinos, which improves the limits by 10\% to 16\%.
The value of the average upper limit presented in this work corresponds to twice the limit on 
the muon neutrino flux $\mathrm{\Phi_{\nu_\mu+\bar{\nu}_\mu}^\mathrm{0}}
= 5.5 \cdot 10^{-11} \,\mathrm{TeV}^{-1}\,\mathrm{cm}^{-2}\,\mathrm{s}^{-1}$.  
A stacking analysis for preselected active galactic nuclei and a
search based on the angular separation of the events were also performed.
We report the most stringent flux upper limits to date,
including the results of a detailed assessment of systematic
uncertainties.

\end{abstract}

     
\maketitle

\section{Introduction}
\label{Sec:Intro}
The search for high energy extraterrestrial neutrinos is the major
focus of the Antarctic Muon And Neutrino
Detector Array (AMANDA)~\cite{andres2000}. The goal is the understanding of the
origin of high energy cosmic rays. 
While a flux of 
charged particles is observed up to energies of a
few hundred EeV, high energy gamma rays with energies up to a few tens
TeV have been detected from several astrophysical objects.  
Remarkably, the nature of the high energy
processes leading to the observed particles and radiation is in most
cases not known.

Neutrinos are expected to be emitted from a variety of 
astrophysical objects: galactic objects like pulsars~\cite{link2005}, accreting binary
systems~\cite{anchordoqui2003}, particularly
micro-quasars~\cite{aharonian2006,torres2006,distefano2002},
and supernova remnants~\cite{bednarek2005b}, as well as from 
extragalactic
objects like active galactic nuclei~\cite{stecker2005,alvarez2004}, particularly
blazars~\cite{mannheim2001,muecke2001,muecke2003,neronov2002b}. Reviews that
include flux
predictions of high energy neutrinos from galactic and
extragalactic objects can be found 
in~\cite{bednarek2005b} and~\cite{learned2000}. 
To date no extraterrestrial high energy neutrino flux has been
observed~\cite{ahrens1997:diffuse,ackermann2004:cascade,aynutdinov,rhode1996,ambrosio2003}. 
Searches for point sources of high energy neutrinos were
presented by~\cite{abe2006,ambrosio2001,svoboda1995,adarkar1991,oyama}.
 
The search for cosmic neutrinos appears more challenging than the
observation of cosmic rays and high energy gamma-rays, due to the much
smaller cross section for neutrino interaction.
On the other hand, 
the small interaction cross section 
makes neutrinos rather unique astronomical
messengers: neutrinos point back to their origin and unlike gamma-rays
they can escape from dense matter regions and
propagate freely over cosmological distances. Their observation would
provide an incontrovertible signature of hadron
acceleration by astrophysical objects.

In this paper we report the results of a search for point sources of  
high energy neutrinos using data collected with AMANDA-II between 2000
and 2004.

\section{Detection of upward going neutrinos with AMANDA} 
\label{Sec:Detect}
The AMANDA-II detector is located at the Geographic South Pole and 
consists of an array
of photomultipliers to detect \cere photons emitted by charged particles
traversing the polar ice.    
An individual detection unit (optical module) is assembled from
an 8 inch diameter photomultiplier, providing good sensitivity 
to single photons, housed in a pressure-resistant glass sphere, both optically 
coupled with transparent gel. The system has been mechanically and optically
stable since the first year of deployment (1996).
Completed in the year 2000, the detector  
includes 677 optical modules 
on 19 vertical strings, most of which are deployed at 
depths between 1.5 and 2 kilometers~\cite{andres2000}. Approximately
540 of the optical modules that form the core of the detector array 
and showing stable performance are used for this analysis.

The geometry of 
AMANDA-II is optimized to detect muon tracks induced by charged current
interactions of neutrinos with energies above 1 TeV. 
Neutrino induced
muon tracks may have ranges of several kilometers (about 8 km in ice
at 10 TeV). They 
are reconstructed from the arrival time of the
\cere photons at the optical modules.  
The energy threshold depends on reconstruction methods and quality
criteria. In this analysis 95\% of the Monte Carlo simulated
atmospheric neutrinos have energies larger than about
100 GeV (\cite{lipari1993,honda2004}).
The muon energy can be estimated 
from the number of detected \cere photons.
The resolution in the logarithm of the energy, $\mathrm{log}_{10}$(E/GeV), is 
about 0.4 at energies above a few TeV~\cite{ahrens2004:reco}.
Above 1 TeV, the mean angular offset between the incoming neutrino and
the muon track  
is less than $0.8^{\circ}$~\cite{gazizov2005}. The
mean scattering angle due to multiple Coulomb-scattering during propagation
of the muon is an order of magnitude smaller~\cite{hagiwara}.

Searches for astrophysical sources of neutrinos have
to cope with a background of events from
the interaction of cosmic rays in the earth's atmosphere. 
Decays of secondary mesons induce a background of downward going muons
and a more uniform background of neutrinos.  
Typical trigger rates measured with AMANDA-II
are $O(10^9)$ events per year from downward going
atmospheric muons and  $O(10^3)$ muon tracks induced by
atmospheric neutrinos, while only a few events are predicted by
models for astrophysical sources~\cite{gaisser1995,halzen2002}. 
Neutrino candidates are selected by rejecting muon tracks reconstructed as
downward going since only neutrinos can cross the earth. This limits
the sensitivity to the northern sky.

A point source would manifest itself
as a localized excess of events over the background. While the background is
uniformly distributed in right ascension, 
the angular distribution of an astrophysical signal would 
follow the detector point spread function.  In order to achieve
a high signal-to-noise ratio, much effort was
 dedicated to improving the event reconstruction 
and selection, and consequently the track angular resolution, 
over a wide energy range. 

A series of reconstruction methods with increasing accuracy at the
expense of increased reconstruction time are applied. 
Fast pattern recognition procedures  
provide a first-guess estimate of the track direction. 
Because of the scattering of the photons on dust and 
crystal grains in the polar ice~\cite{woschnagg2005}, 
complex reconstruction algorithms are necessary to
measure the direction with a good angular resolution. 
Based on maximum likelihood 
procedures in a multi-parameter space
using the first-guess results as starting point,
high level reconstructions aim
at finding  
the best likelihood for a given event topology with respect
to the recorded hits~\cite{ahrens2004:reco}.

About 0.1\% of the downward going muons  
are wrongly reconstructed
as upward going. 
A selection based on event quality parameters is used to reduce these events
by an additional four orders of magnitude.
Yet an irreducible background remains from
upward going muons induced by atmospheric neutrinos 
together with a small fraction of mis-reconstructed downward going muons
plus possible signal events. 
Typical resolutions achieved in the reconstruction of the muon direction
are between  $1.5^{\circ}$ and
$2.5^{\circ}$ degrees (median spatial angle), 
depending on energy and declination. 

In order to avoid 
biases in the event selection, the final event selection was
developed following a blind approach. In the search for point sources, where
the event direction is used to look for a signal,
this is accomplished by optimizing cuts on a sample of events with
randomized right ascension.
Accumulations of
events due to signal would be averaged out, while the
dependency of the detection efficiency on declination is preserved. 
The background 
is estimated from the detected events, by adopting a 
technique similar to the ``off-source'' method in gamma-ray astronomy. 
The error of the background estimation is therefore small and statistical
only, independent of
the detector simulation.

The detection efficiency for astrophysical neutrinos is studied with
a complete Monte Carlo description of neutrinos fluxes, propagation
through the earth and interactions, of the muon propagation and of the 
detector response~\cite{ahrens1997:point}. 
The latter takes into account the propagation of photons
in the ice and the photon detection probability. The systematic uncertainties
in this modeling affect the signal efficiency and therefore the
calculation of flux upper limits or, in case of detection, the
precision with which the cosmic
neutrino flux can be measured. Comparison of the final event sample 
 to the Monte Carlo expectation for atmospheric
neutrinos allows the verification of the modeling
accuracy and of the detection 
efficiency. These aspects will be addressed in detail in
Section~\ref{Sec:Sys}.

\section{Event reconstruction and selection}
\label{Sec:Rec}
The searches reported in this paper use the data collected with the
AMANDA-II detector in the years 2000 to 2004. The austral-summer 
data (from November to February), 
taken during the detector maintenance and station summer activity 
periods, are excluded. Periods of overall detector
instability are also discarded. The remaining live-time is 1001
days, after correction for the intrinsic DAQ dead-time.
The trigger used to collect this data
  requires at least 24 optical modules (OM) recording one or more pulses
above threshold (hits) within 2.5 $\mu$s.

Table~\ref{Tab:Proc} shows the first three filtering levels used to process
the 8.9 $\times 10^9$ events used in this analysis. The multi-level
filtering is needed because the final reconstruction algorithms are
too CPU-intensive to use on the entire dataset. 
Sophistication and CPU demand per event of these procedures increase with 
level, as does the tightness of cuts for background rejection.
The event passing rates in Table~\ref{Tab:Proc} are normalized to the number of triggered
events (8.9 $\times 10^9$). {\it Level 1} and {\it Level 2} of the event
reconstruction and selection are based on relatively loose cuts,
in order to extract an event sample which is still useful for other analyses.

Details of the pre-processing techniques (hit and optical module
selection) and of the reconstruction algorithms can be found
in~\cite{ahrens2004:reco}. Before reconstruction, short pulses 
are removed 
which can be ascribed to electronic noise.
Hits from unstable optical modules are also rejected based on their
typical TDC and dark noise rates compared to the average 
(hit and optical module selection in
Table~\ref{Tab:Proc}). 
Events are required to have at 
least 24 modules hit after this cleaning as in the
hardware trigger (re-trigger in Table~\ref{Tab:Proc}). 

Two fast pattern recognition
algorithms are then applied to reconstruct the direction of
the muons: DirectWalk, described in~\cite{ahrens2004:reco}, and JAMS. JAMS provides an enhanced
downward going muon track rejection power compared to previous
results~\cite{ahrens2004:point,ackermann2005:point}. 
The best guess for the direction of a muon track 
is found from the distribution of hits projected 
on a plane orthogonal to a candidate track direction. 
Only hits with a short delay compared 
to the arrival time expected for the direction of the  
track hypothesis are considered. 
Photons generating such  ``direct'' hits have undergone
only a few scatters in the ice
and have therefore preserved the 
directional information. The track direction hypothesis
is then varied and the distribution of the hit projections studied. 
The direction with
the largest and most isotropic cluster of associated hits is chosen as
JAMS result. 

\renewcommand{\arraystretch}{0.7}
\begin{table}[!tbh]
 \begin{center}
  \begin{tabular}{c|l|l|l|e{2.2}}
Level &Hit/Event filter    &Track reconstruction &Event cut&\talignc{Events kept} \\ \hline
1     &Hit \& OM           & & & \\
      &Re-trigger          &                     &hit multiplicity$>$23 &95.0\% \\
      &                    &DirectWalk           &${\theta}_{\mathrm{DW}}>$70$^{\circ}$ &3.7\% \\
\hline
2     &                    &JAMS                 &${\theta}_{\mathrm{JAMS}}>$80$^{\circ}$ &0.4\% \\
      & Cross-talk         & & & \\
\hline
3     &                    &Unbiased likelihood fit (UL)
                           &${\theta}_{\mathrm{UL}}>$80$^{\circ}$   &0.1\% \\
      &                    & Bayesian likelihood fit (BL)   & & \\ 
 \end{tabular}
 \caption{\label{Tab:Proc}Summary of the reconstruction and filtering
    steps as explained in the text for the first three levels of
    data reduction, with the fraction of events passing each level
compared to the number of triggered events (8.9 $\times 10^9$). }
 \end{center}
 \end{table}
\renewcommand{\arraystretch}{1}

With JAMS we are able to reject classes of downward going muons which 
the DirectWalk fit wrongly reconstructs as upward going particles.
As a consequence, an  
efficient reduction of the background from atmospheric muons by a
factor of 250 can be
achieved at {\it Level 2} of the event selection, applying 
angular cuts to the directions from both first-guess
algorithms (cfr. Table~\ref{Tab:Proc}).
A filter based on the amplitude and duration of hits and on a
talker-receiver map is then applied to exclude pulses induced along 
twisted-pair cables when analog signals are transmitted from optical
modules to the surface (cross-talk in Table~\ref{Tab:Proc}).

Two iterative reconstructions follow:
an unbiased likelihood fit (UL), seeded with the result of JAMS and with
32 randomly chosen input directions, and a Bayesian likelihood 
fit (BL),  
seeded with the results of UL and with
64 randomly chosen input directions.
The Bayesian fit incorporates a 
prior hypothesis with a parameterization of the MC zenith 
distribution for atmospheric muons at the detector~\cite{ahrens2004:reco}.
The final direction is defined by the best likelihood found.

At {\it Level 3} the data sample is reduced to 9.9 $\times 10^6$ tracks and is
still dominated by 
downward going muons, outnumbering neutrinos by three orders of magnitude. 
Fake events due to non-simulated
electronic artifacts are rejected after {\it Level 3} with 
a filter sensitive to correlated noise~\cite{pohl:phd}. The event
reconstruction and selection has proved to be stable with respect to these
detector instabilities.

Neutrino induced upward going tracks are selected after {\it Level 3}
by imposing event quality requirements based on the single track angular
resolution and on  
topological parameters describing the distribution of hits along 
the trajectories. Three independent
parameters are chosen: a) the event based angular resolution,  
proportional to
the width of the likelihood minimum and derived from the fit error 
matrix~\cite{neunhoeffer2006a}, b) the smoothness,
a parameter describing the homogeneity of the hits along the
track~\cite{ahrens2004:reco}, and c) the ratio of the likelihoods from the
unbiased and the Bayesian reconstructions. 

Distributions of these
observables were constructed for both data and 
signal Monte Carlo in 22 declination bands. Together with the search
bin radius of the binned search defined in Section \ref{Sec:SteadyPoint}, the 
parameter space of these variables is scanned to find 
the optimum selection with respect to 
signal efficiency and residual background. 
The optimum selection provides the best sensitivity
as the average upper limit in absence of a signal~\cite{hill2003a,feldman1998}.

The optimum selection criteria determined with this method depend
on the assumed signal light deposited and therefore on the assumed
signal energy spectrum and on the track direction.
We implemented event cut optimizations assuming different
signal energy power-law spectra
$\frac{d\Phi}{dE}={\Phi}^\mathrm{0}\cdot(E/\mathrm{1\,TeV})^{-\gamma}$, with
${\Phi}^\mathrm{0}$ as normalization. 
Two spectral indices were considered: $\gamma$=2, generally
assumed to be the most likely for astrophysical beam dumps, following 
Fermi shock acceleration of protons, and $\gamma$=3
as a possible extreme of softer spectrum 
scenarios\footnote{A $\gamma$=2 spectrum
with a 1 TeV cutoff was also considered, however the optimum selection found is
identical to the $\gamma$=3 case. Therefore this case is omitted
here.}. We chose cut values which are 
close to the individual optima and
provide a good sensitivity in both cases.

The optimum size of the circular search bins varies between 2.25$^{\circ}$ and
3.75$^{\circ}$ depending on declination. 
These search bins contain 60\% to 80\% of the simulated signal, respectively.

\section{Properties of the final event sample}
\label{Sec:Pro}
A final sample of 4282 upward going muon-like events survived the cuts.
This is in agreement with expectations from a
Monte Carlo simulation of atmospheric 
neutrinos following the parametrization 
in~\cite{lipari1993}. The central value of this
parametrization yields
4600$^{+300}_{-1000}$(sys) expected events. 
The systematic error is discussed in Section~\ref{Sec:Sys}.
 
\begin{figure}[!t]
\begin{center}
\includegraphics[width=4in]{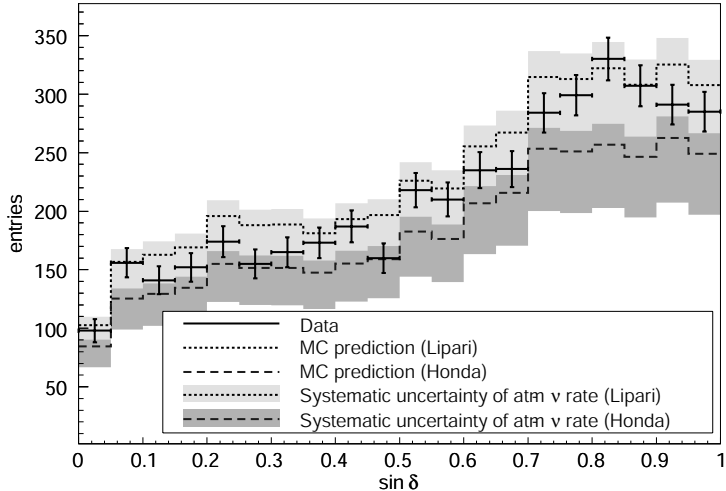}
\end{center}
\vskip -0.6cm
\caption{Declination angle distribution of the final event selection 
compared to the
expectation from Monte Carlo simulation of atmospheric neutrinos,
including the systematic error band (see Section~\ref{Sec:Sys}). 
 The two extremes~\cite{lipari1993,honda2004} among different
predictions are shown. Error bars on the data point are statistical.}
\label{Fig:Zenith}
\end{figure}
We estimate the contamination from
mis-reconstructed downward going 
events to be less than 5\%.
This is obtained from the comparison of the event sample after 
{\it Level 3} of the data reduction to the prediction from atmospheric
neutrinos, as a function of the quality of the
reconstructed tracks~\cite{ahrens2002:atmospheric}.
 
Figure~\ref{Fig:Zenith} compares the observed declination
distribution to the one expected for 
atmospheric neutrinos. Simulation results are given for 
two different parameterizations of atmospheric neutrino 
fluxes~\cite{lipari1993,honda2004}. The systematic errors are
indicated by shadowed areas (see Section~\ref{Sec:Sys}).
The angular distribution confirms 
that the background for sources other 
than atmospheric neutrinos is small, within the model uncertainties.

\begin{figure}[!thb]
\begin{center}
\includegraphics[width=4in]{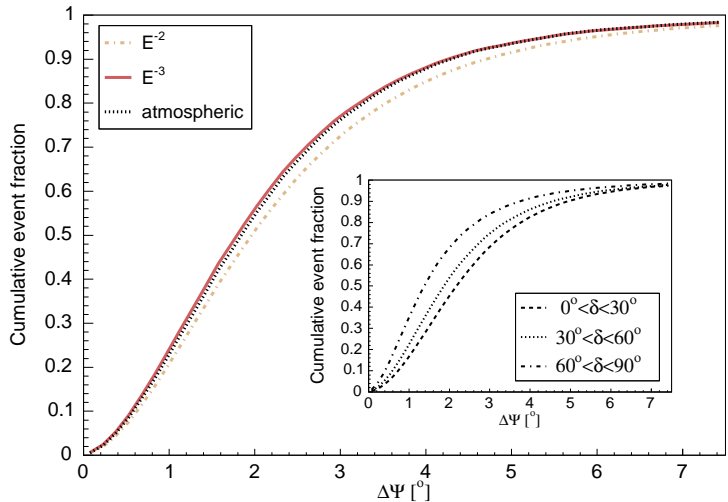}
\end{center}
\vskip -0.6cm
\caption{Cumulative point spread function for the events passing the
selection, for different simulated energy spectra. The inset graph
shows the dependence on declination, for a spectral index $\gamma$=2.
$\Delta\Psi$ is the space angle difference between the true and the
reconstructed direction of simulated events.}
\label{Fig:Psf}
\end{figure}

The average pointing resolution of the selected
events can be estimated from the distribution of the directional
difference between simulated neutrino tracks and reconstructed muon
tracks of Monte Carlo events. 
For a spectral index $\gamma$=2, the median value of the space
angle distribution is typically 2$^\circ$. 
Figure~\ref{Fig:Psf} shows the cumulative point spread
function after the final event cuts for different energy spectra. It
can be seen that the angular resolution is declination dependent.

\begin{figure}[thb]
\begin{center}
\includegraphics[width=6.5in]{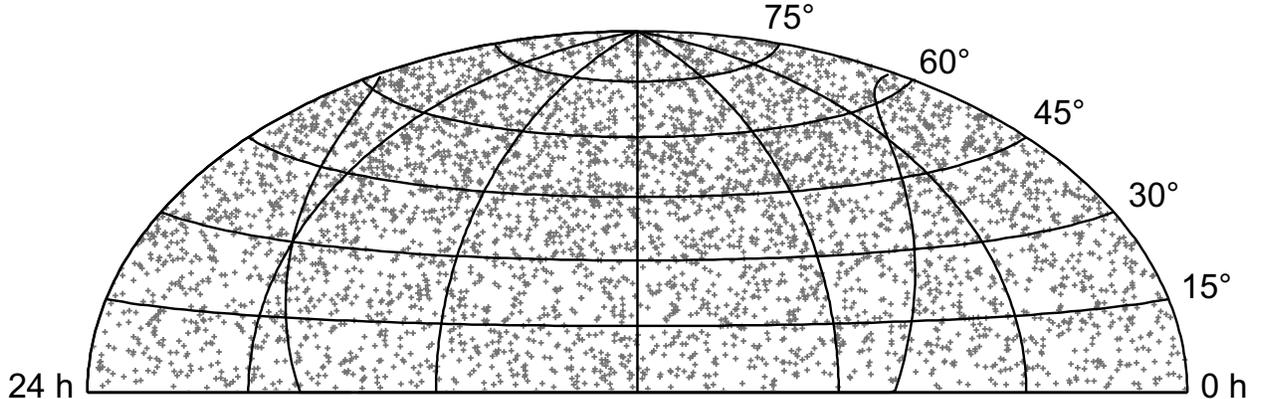}
\end{center}
\vskip -0.6cm
\caption{Sky-plot of the selected 4282
upward going neutrino candidate events. Horizontal coordinates are
given as right
ascension and vertical coordinates as declination. The black line marks the 
location of the galactic plane.}
\label{Fig:Sky}
\end{figure}

The directions of the selected upward going events are shown in
Fig.~\ref{Fig:Sky}. Our event selection 
yields a relatively uniform coverage of the 
northern sky. Moreover, the polar location of AMANDA assures an equal
exposure for all declinations, independent of the detector operation periods.

This analysis is primarily sensitive to events induced by 
muon neutrino charged current interactions. Tau neutrinos give an
additional contribution via
charged current interactions
followed by the 
$\tau^{\pm} \rightarrow \mu^{\pm} + \overline{\nu}_{\mu}(\nu_{\mu}) + {\nu}_{\tau}(\overline{\nu}_{\tau})$ 
decay, with a 17.7\% branching
ratio~\cite{pdg}, which is
included in the upper limits reported in Section~\ref{Sec:SteadyPoint}.
To estimate the tau neutrino contribution to the final event
sample, tau neutrinos were generated 
according to~\cite{gazizov2005} and propagated through the standard AMANDA-II
simulation chain. 

Under the assumption of equal fluxes of cosmic muon and tau neutrinos
at the earth, the additional contribution of tau neutrino signal events 
ranges from 10\% to 16\% for $\gamma$=2, depending on declination.
This assumption is in accordance with the generally assumed scenario 
of a flavor ratio at the earth of 
${\Phi}_{{\nu}_e}$~:~${\Phi}_{{\nu}_\mu}$~:~${\Phi}_{{\nu}_\tau}$
= $1 : 1 : 1$, after neutrino oscillation. 
Deviations from this case can emerge at high energies,
where in some astrophysical
scenarios the contribution to the neutrino flux 
from muon decay is
suppressed~\cite{kashti2005}, leading to
${\Phi}_{{\nu}_e}$~:~${\Phi}_{{\nu}_\mu}$~:~${\Phi}_{{\nu}_\tau}$
= $1 : 1.8 : 1.8$. However, 
equal muon and tau neutrino fluxes are still expected 
in this scenario.

Two-flavor oscillations of atmospheric neutrinos were simulated with
$\Delta {m_{2,3}}^2 = 2.5 \times 10^{-3} \, \mathrm{eV}^2$ and maximum mixing
$\theta_{2,3}=45^{\circ}$~\cite{ashie2005}. 
For our sensitive energy range, this results in
a disappearance of muon neutrinos of less than 3\%, depending on the
direction. The corresponding appearance of tau neutrinos leads to an
increase of the detected muon rate which is less than 0.5\% and is neglected for this
analysis.

\begin{figure}[!thb]
\begin{center}
\includegraphics[width=3.3in]{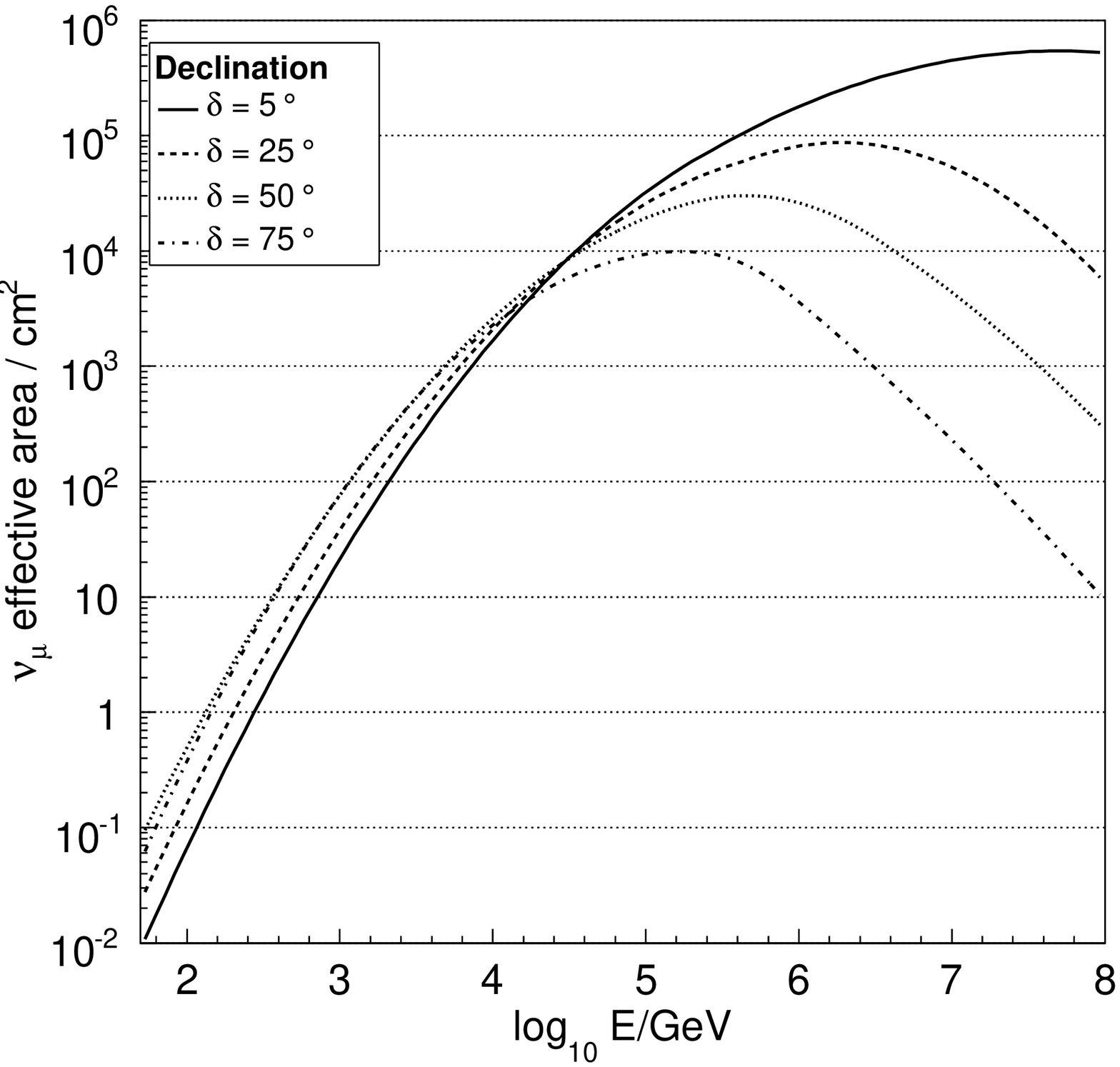}\includegraphics[width=3.3in]{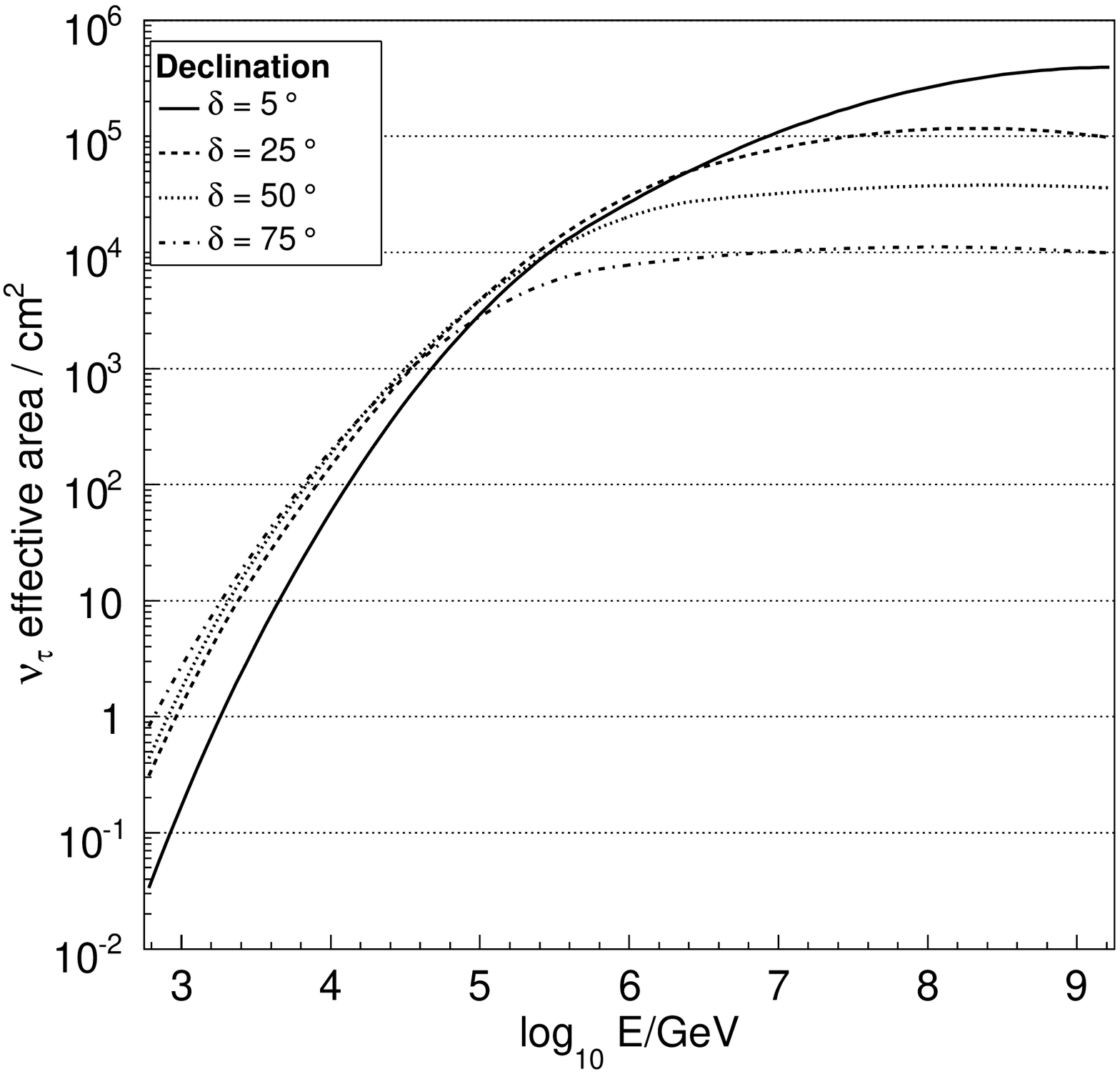}
\end{center}
\vskip -0.6cm
\caption{Neutrino effective area as a function of the neutrino
energy for different declinations for muon neutrinos (left) and tau
neutrinos (right) respectively. The decrease observed at high energies
on the left is due to neutrino absorption in the earth. }
\label{Fig:Acc}
\end{figure}

The neutrino effective area is a convolution of the neutrino interaction cross
section, the muon
survival probability and the detector response (geometry and
detection efficiencies). It depends on the neutrino energy and
direction
as shown in Figure~\ref{Fig:Acc} for muon neutrinos (left) and tau
neutrinos (right), including the earth shadowing effect.

Because the event selection optimization allows wider spectral
scenarios than in our previous point source 
searches~\cite{ahrens2004:point,ackermann2005:point}, the sample of selected up-going events
contains
a significant contribution from low energy events ($E_{\nu}<$1 TeV).
In a Monte Carlo simulation of atmospheric neutrinos~\cite{lipari1993}, 95\% of the
events have energies larger than about 100 GeV. The median energy is about 500 GeV.
Neutrinos from a source with a $\gamma$=2 spectrum, however, are 
expected to carry a median energy of about 50 TeV, due to the
cross-section, the muon range and the detection and event selection
efficiency, which leads to a steeply rising sensitivity with energy.
Table~\ref{Tab:Ene}
shows relevant information on the energy distribution 
of Monte Carlo events, for different input spectra.
\renewcommand{\arraystretch}{0.7}
\begin{table}[!bth]
 \begin{center}
  \begin{tabular}{e{2}|e{1.1}e{2.2}c|e{2.2}e{2.2}c|e{2.2}}
\talignc{$f(E_{\nu}>E_{th})$} 
&\multicolumn{3}{c}{$\gamma$=2} 
&\multicolumn{3}{c}{$\gamma$=3}  
&\talignc{Atm.} \\ \hline
     &\talignc{$\nu_\mu$} &\talignc{$\nu_\tau$} &$\nu_\mu+\nu_\tau$
     &\talignc{$\nu_\mu$} &\talignc{$\nu_\tau$} &$\nu_\mu+\nu_\tau$
     &\talignc{$\nu_\mu$} \\
95\% &3.2 &3.6 &3.2 &2.1 &2.5 &2.1 &2.0 \\
50\% &4.7 &5.2 &4.7 &3.0 &3.3 &3.0 &2.8 \\
 5\% &6.2 &6.9 &6.4 &4.4 &4.7 &4.4 &3.9 \\
 \end{tabular}
  \caption{\label{Tab:Ene}
Energy values ($E_{th}$) above which a fraction of
neutrino (and anti-neutrino) events $f(E_{\nu}>E_{th})$ is observed
integrated in declination,
for different input spectra. $E_{th}$ is given in
$\mathrm{log}_{10}$($E_{th}$/GeV). In case of atmospheric neutrinos no
contribution from tau neutrinos is expected in the final data sample
(see text for discussion).}
 \end{center}
  \end{table}
\renewcommand{\arraystretch}{1}

The limit setting capability of this analysis can be expressed by the 
sensitivity to neutrino fluxes from point sources, introduced in
Section~\ref{Sec:Rec}. The sensitivity is a function of the
background and describes the observation potential. In case no excess
is detected over the expected background, we calculate the upper
limits to the neutrino flux, as a function of both background and
experimental observations. In this work we give both the sensitivity and
upper limits to
the parameter ${\Phi}^\mathrm{0}$, i.e. on the normalization constant
to the differential flux 
$\frac{d\Phi}{dE}=\mathrm{\Phi^{0}\cdot\big(\frac{E}{1\,TeV}\big)^{-\gamma}}$.

\begin{figure}[!tb]
\begin{center}
\includegraphics[width=4in]{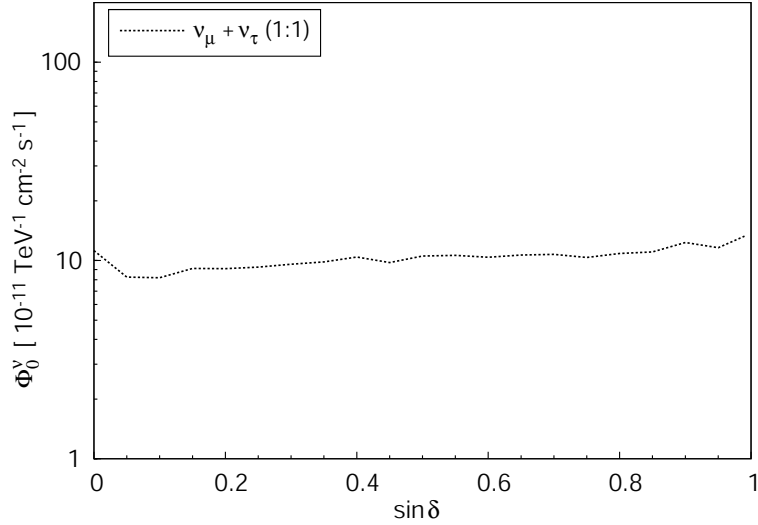}
\end{center}
\vskip -0.6cm
\caption{Sensitivity to point sources of neutrinos 
as a function of declination, for an integrated exposure of 1001 days
and 
for a $\gamma$=2 source spectrum.
The sum of muon and tau neutrinos fluxes is considered assuming
$\mathrm{\Phi_{\nu_\mu+\bar{\nu}_\mu}/\Phi_{\nu_\tau+\bar{\nu}_\tau}}$=1
}
\label{Fig:Sens}
\end{figure}

The sensitivity of this analysis to point sources of muon and tau neutrinos (and anti-neutrinos)
with a spectral index $\gamma$=2 and
averaged over declination 
is
$\mathrm{\Phi_{\nu_\mu+\bar{\nu}_\mu}^\mathrm{0}+\Phi_{\nu_\tau+\bar{\nu}_\tau}^\mathrm{0}}
= 10.0
\cdot 10^{-11} \,\mathrm{TeV}^{-1}\,\mathrm{cm}^{-2}\,\mathrm{s}^{-1}$,
assuming a flavor ratio at the earth 
$\mathrm{\Phi_{\nu_\mu+\bar{\nu}_\mu}/\Phi_{\nu_\tau+\bar{\nu}_\tau}}$=1
(Figure~\ref{Fig:Sens}).

\section{Systematic errors}
\label{Sec:Sys}
\subsection{Overview}
Three main classes of systematic errors affect the searches reported
in this work: 
the uncertainty in the optical module response (timing
resolution and optical module sensitivity),
the uncertainty in the modeling of the
neutrino and muon propagation and interaction, 
and other simplifications in the simulation (propagation of photons in the
ice, detector response and neutrino-muon scattering angle).

We estimate the systematic error on the rate of high energy neutrinos
by variations of these quantities in the input to the simulation. The
results typically depend on neutrino energy and, therefore, 
on the assumptions on the cosmic neutrino energy spectrum. 
The dominant error is due to the uncertainty
in the optical module sensitivity. 
For a spectral index $\gamma$=2 it contributes $^{+2}_{-9}$\%
of the total systematic error
of $^{+10}_{-15}$\%~\cite{ackermann:phd}.

Table~\ref{Tab:Sys} summarizes the systematic errors on the rate of high energy neutrinos
estimated for this analysis, for three assumed energy
spectra: $\gamma=2$, $\gamma=3$ and atmospheric energy spectrum,
according to~\cite{lipari1993}.

\renewcommand{\arraystretch}{0.7}
\begin{table}[htb]
 \begin{center}
  \begin{tabular}{c|l|ccc}
Class 
&\talignc{Source of uncertainty}
&\talignc{$E^{-2}$}&\talignc{$E^{-3}$}          
& Atm. \\ \hline

1 &Optical module timing resolution &$\pm$2\% &$\pm$2\% &$\pm$2\%\\ 
  &Optical module sensitivity &$^{+2}_{-9}$\% &$^{+5}_{-17}$\% &$^{+6}_{-19}$\%
\\ \hline
2 &Neutrino cross section and rock density &$\pm$8\% &$\pm$3\% &$\pm$3\%\\
  &Muon energy loss &$\pm$1\% &$\pm$1\% &$\pm$1\%
\\ \hline
3 &Photon propagation in ice &$\pm$5\% &$\pm$5\% &$\pm$5\%\\
  &Reconstruction bias & $^{+0}_{-7}$\% &$^{+0}_{-8}$\% &$^{+0}_{-9}$\%\\ 
  &Neutrino-muon scattering angle &$^{+0}_{-1}$\% &$^{+0}_{-8}$\% &$^{+0}_{-13}$\% \\\hline
& {\bf Sum}                 
& ${\bf ^{+10}_{-15}}${\bf \%} 
& ${\bf ^{+6}_{-21}}${\bf \%}
& ${\bf ^{+7}_{-25}}${\bf \%}\\
 \end{tabular}
  \caption{\label{Tab:Sys}
Summary of the systematic error in the measured rate of high
energy muon neutrinos due to the three classes of systematic uncertainties, 
for different assumption on the energy spectrum.}
 \end{center}
  \end{table}
\renewcommand{\arraystretch}{1}

An independent confirmation of the estimate of the systematic errors
summarized in  Table~\ref{Tab:Sys} is given by the agreement of the absolute flux
prediction of atmospheric neutrinos to the experimental results.
Figure~\ref{Fig:Zenith} compares the
zenith angle distribution for the observed events and the Monte Carlo
simulation for atmospheric neutrinos, including the systematic error
band. This confirmation is however limited by the 
theoretical uncertainty on the absolute flux prediction
of atmospheric neutrinos (up to 25\% or more at the highest
energies~\cite{gaisser2002}). 
The extremes of these 
predictions are shown in Figure~\ref{Fig:Zenith}~\cite{lipari1993,honda2004},
the difference representing the theoretical uncertainty
due to the limited knowledge of the total primary cosmic ray flux and
of the hadronic interactions in the atmosphere. It can be seen that
the observed angular distribution and the event rate range between the two extremes.
The theoretical uncertainty on the atmospheric neutrino 
flux prediction does not affect our searches for point sources, as
the background is measured from the data.

The total systematic error in the neutrino rate (the sum in Table~\ref{Tab:Sys}) 
is asymmetric, while in the upper limit
calculation errors are assumed to be Gaussian. In the limits reported in
section~\ref{Sec:SteadyPoint} we therefore round up
the total error on the neutrino rate to 15\% for $\gamma=2$ and 20\% for $\gamma=3$. 
The main classes of systematic uncertainties are discussed in detail
in the next sections\footnote{In order 
to apply a symmetric systematic error on the limits a shift of 2.5\% (7\%) was also
applied to the signal prediction, respectively for $\gamma$=2~(3).}. 

\subsection{Optical module response}
Two sources of uncertainty are related to the response of the optical
modules: 
the timing resolution and the optical module sensitivity. The first can influence the accuracy of the track
reconstruction while the second affects the effective area.

The timing of the optical modules, measured using YAG laser pulses,
is better than 5 ns~\cite{andres2000}. The impact on the rate of
selected upward going events of this intrinsic resolution is less
than 2\%, independent of the spectral index. 
This uncertainty is also checked by comparing the measured and
the expected arrival time of photons for downward going muon tracks. 

The optical module sensitivity
depends on the photomultiplier
quantum efficiency, on the transmission
properties of the glass sphere coupled to 
the optical gel and on the propagation of photons in the local
(re-frozen) ice surrounding the optical modules. 
A nominal value of the optical module sensitivity is an input parameter
in the detector simulation.

A conservative overall
uncertainty on the optical module sensitivity of 30\% or more
(including the uncertainties on the properties of the re-frozen ice) was 
used in previous works 
(e.g.\cite{ahrens1997:point,ahrens2003:cascade,ackermann2004:cascade}). This was 
dominated by the uncertainty on the optical gel
transmittance based on laboratory measurements performed 
on a series of spare samples.
This uncertainty can however be largely reduced by comparison of the
measured and simulated zenith angle distribution of the selected
upward going events. Due to the inhomogeneous distribution of the optical
modules in the detector -- arranged along strings in a cylindrical
structure with larger vertical dimension compared to the 
horizontal ones -- reconstructed muon 
tracks are differently affected by the optical module sensitivity
according to their direction. 
We studied the effect varying the
nominal value of the optical module sensitivity in the Monte Carlo
simulation within a range of $\pm$ 30\%. 
Comparing 
the angular distribution of
simulated and observed events, 
we found a 1$\sigma$ range of the best fit sensitivity
corresponding to
$100^{+3}_{-10}\%$ of the nominal value used in the detector
simulation.
The impact on the rate of high energy neutrino events 
is $^{+2}_{-9}$\% for $\gamma$=2 (Table~\ref{Tab:Sys}).
This result is
stable with respect to the other sources of
systematic uncertainty. 
This estimate also includes
the systematic error due to the uncertainty of the absolute
light yield from secondary cascades along the muon track.

\subsection{Neutrino and muon propagation and interaction}
The number of high energy muons passing through the AMANDA-II detector
for a given flux of neutrinos depends on the neutrino absorption in
the earth and on the rate of neutrino interactions in the
column of ice and bedrock surrounding the detector and acting as
target for the production of detectable muons. Limited geophysical
measurements are available to determine the rock density at the
South Pole. Typical rock samples are found to vary by 10\% around the
nominal density of 2650
kg/$\mbox{m}^3$~\cite{lowry1997} used in the Monte Carlo simulation for this
analysis. This uncertainty alters the signal prediction for a neutrino flux
with a spectral index of $\gamma$=2
by 2\% for nearly horizontal events ($\delta=0^{\circ}$ to $30^{\circ}$)
and 7\% for nearly vertical events ($\delta=60^{\circ}$ to $90^{\circ}$). 
For softer spectra this uncertainty has a negligible effect. 
Most detectable muons are produced in
the ice surrounding the detector, which has a well known density,
rather than in the bedrock. 
This error includes the impact of this uncertainty on the muon
propagation.

An uncertainty of 3\% is estimated for the charged current
deep-inelastic neutrino-nucleon scattering and therefore for the
corresponding muon
event rates. This uncertainty 
is estimated from the error table on the
Parton Density Functions in the range between 100 GeV and 1 PeV, 
as reported in~\cite{cteq:www} and the
prescription for the error analysis in~\cite{pumplin2002}. 

For a spectral index $\gamma$=2, these two
sources of uncertainty together give an overall error of
4\% (8\%) for horizontal (vertical) events. 

The rate of muons crossing the detector also depends on the muon
energy loss. The processes considered in the Monte
Carlo simulation for the energy loss and the production of secondaries
are ionization, bremsstrahlung, pair production, photo-nuclear
interaction and decay, known within an uncertainty 
between 1\% for muon energies of about 1TeV and a few percent for
higher energies~\cite{bugaev2000}. A higher uncertainty can affect
the tau energy loss at high energies~\cite{bugaev2004}. However, due
to the short lifetime of taus, the impact of this error on the rate
of detected events is expected to be smaller than the case of muon tracks. 
The resulting systematic error of the absolute event
rate is 1\% for a variation of the cross sections by~$\pm$~2\%.

\subsection{Other simplifications in the simulation}
The glacial ice contains impurities which reflect past 
climatological changes. Calibration light sources 
were used to map the absorption and the scattering length of
the ice and develop a wavelength and depth-dependent model of
its optical properties~\cite{woschnagg2005}. A detailed photon
propagation in the simulation is, however, computationally intensive.
Therefore past AMANDA
analyses used a simplified model, averaging the ice properties over the
wavelengths of the \cere spectrum and over the 
depths covered by AMANDA, 
optimized for the best agreement in the timing properties of
observed and simulated downward going tracks. 
A comparison of the results for this analysis using the
``average'' model to more recent developments (i.e. a detailed 
depth-dependent simulation 
based on the results in~\cite{woschnagg2005})
gives a variation in the passing rate of the
final events of 5\% or less.

Residual and non-identified sources of systematic uncertainties
appear
when comparing the observed and the simulated cut parameter distributions
at the final level, assuming that the sample is dominated by atmospheric
neutrinos. A certain level of mismatch is observed for the angular resolution
and the smoothness parameter. This is likely due to inaccuracies in
the simulation of the detector response, e.g. non simulated cross-talk and
noise hits, affecting the quality of the reconstructed tracks. A
small residual fraction of mis-reconstructed downward going muons
might also be present
in the data. This is included in the background estimation and
does not affect the simulation of the signal and the relative
systematic uncertainty. 

A conservative estimate of residual
non-identified sources of systematic uncertainties was done
by 
applying a scaling factor to both the smoothness and the angular
resolution parameters, to obtain a good agreement between data and Monte
Carlo simulation at the final level. 
The scaling factors necessary to obtain a
good agreement in the angular resolution and in the smoothness are 1.1
and 1.07, respectively. The final event sample changes by 9\% or less 
when using the scaled parameters rather than the unscaled (7\%
for $\gamma$=2).

An additional systematic uncertainty 
is finally due to neglecting the 
neutrino-muon scattering angle in the simulation.
In this work a neutrino induced muon is
simulated as collinear to the direction of the parent neutrino,
but the energy dependent average angle might exceed the
size of the search bins~\cite{gaisser1990} at energies below 1 TeV.
For a spectral index $\gamma$=2 the systematic overestimation of the
event rate when neglecting the neutrino-muon angle is less
than 1\%. The tau neutrino signal prediction for
this work uses a neutrino generator which properly
accounts for the neutrino-muon angle~\cite{gazizov2005} and is therefore free
of this systematic error. Table~\ref{Tab:Sys} reports the impact of
this uncertainty on the final event rates also for 
the case of atmospheric neutrinos, as an example of soft
spectrum. This uncertainty is however not included in in
Figure~\ref{Fig:Zenith}, since it only affects searches for point sources.

\section{Search for individual point sources in the northern sky}
\label{Sec:SteadyPoint}
A search for point sources of neutrinos is performed with the sample
of 4282 upward going events by looking for 
excesses of events from the directions of known objects and by
performing a survey of
the full northern sky.
This analysis is sensitive to a point source that would manifest
itself as a statistically significant cluster of events within an
angular bin consistent with the point spread function of the detector. 
Both surveys are based on circular search bins of a size 
optimized together with the event selection as described in
Section~\ref{Sec:Rec}.

\subsection{Selected source candidates}
\label{Sec:32}
First, a sample of 32 neutrino source candidates are tested for
an excess of events. The source list (Table~\ref{Tab:32}) includes
galactic and extragalactic objects which are identified sources of high energy 
gamma-rays located in the field of
view of AMANDA-II.
Any source that accelerates charged hadrons to high energies is also
a possible
source of detectable neutrinos from meson decay: 
high energy hadrons will interact with other
nuclei or the ambient photon fields producing hadronic showers. In
these scenarios, high energy photons and neutrinos are expected to be
produced simultaneously, with correlated rates.

We consider five blazars confirmed as
TeV gamma-ray sources~\cite{mannheim2001,muecke2003}, eight blazars
confirmed as (or with indications of being)
GeV gamma-ray sources~\cite{neronov2002b}, eight galactic X-ray binaries
classified as micro-quasars and one neutron star binary
system~\cite{anchordoqui2003}, 
six galactic supernova remnants and
pulsars. We also consider 
individually selected objects, like an EGRET source with high detected flux
above 100 MeV~\cite{hartman1999}, a TeV gamma-ray
active galactic nucleus (M87), the bright and closest active galactic nucleus
NGC~1275 and
an HEGRA TeV gamma-ray source with indications
of hadronic emission~\cite{aharonian2001,butt2003}.

We estimate the number of expected background events per bin ($n_\mathrm{b}$)
from the event density as a function of declination.
Because of the rotation of the earth, any
azimuth-dependent variation in the event rate is averaged out and 
the background density is expected to be uniform in right ascension. 
We average the event densities over right ascension in declination
bands of width comparable to the search bin radius.
The statistical uncertainty in the background
per search bin (${\sigma}_\mathrm{b}$) 
depends on the declination and the largest
value is 7\%.
The background estimate is not affected by systematic uncertainties,
since it is independent of the Monte Carlo description of the
detection efficiency. 
 
\begin{figure}[!t]
\begin{center}
\includegraphics[width=4in]{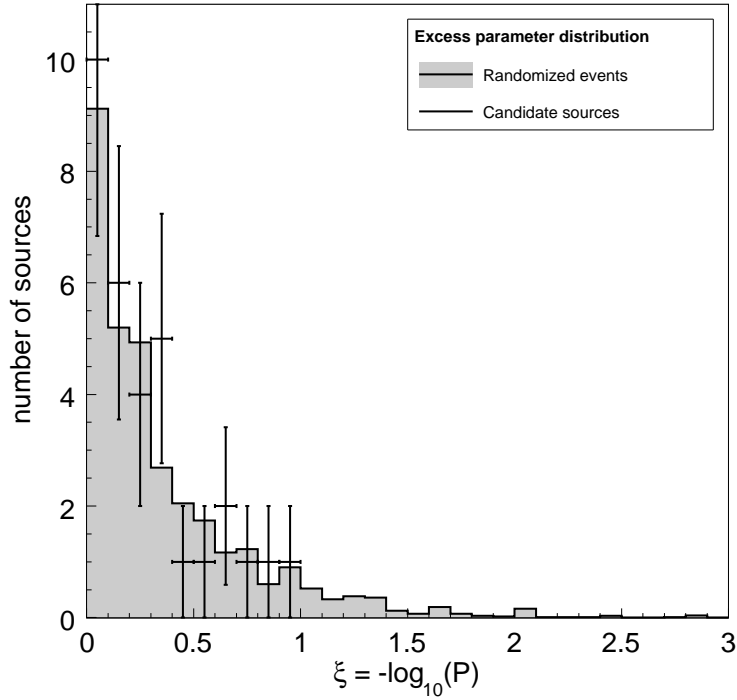}
\end{center}
\vskip -0.6cm
\caption{Excess parameter ($\xi$) distribution from  
the search for neutrinos from pre-selected objects. The
results of each individual observation (data points) are compared to 
the expected
distribution from 1000 simulated equivalent experiments with randomized right
ascension (filled histogram).}
\label{Fig:ExcessPar}
\end{figure}

For each search bin, the number of observed events $n_\mathrm{obs}$ is
compared to $n_\mathrm{b}$. 
If no statistically significant excess 
is found,  we calculate flux upper limits, which depend
on the assumed source location and energy spectrum.
The statistical significance of the 
observations is evaluated with 100 equivalent simulated experiments, 
using the same data sample and randomizing the events in right ascension
independently for
each experiment.
An excess parameter $\xi$ 
is defined as:
\begin{equation}
\xi=-\log_{10}(P)
\end{equation}
where $P$ is the binomial probability of observing $n_\mathrm{obs}$ events in
the search bin, given $N_\mathrm{band}$ events in the declination band used
for the background estimation:
\begin{displaymath}
P(n_\mathrm{obs}|N_\mathrm{band}) = {N_\mathrm{band} \choose n_\mathrm{obs}} \cdot p^{n_\mathrm{obs}}
\cdot {(1-p)}^{N_\mathrm{band}-n_\mathrm{obs}},
\label{Eq:prob}
\end{displaymath}
where $p$ is the probability of a background event in the given search bin.
The parameter $p$ depends 
on the area of the search bin compared to the declination band.
The distribution of the excess parameter for the source candidates agrees with 
the expectation for a purely atmospheric neutrino sample, simulated with
the 100 samples with randomized right ascension (Figure~\ref{Fig:ExcessPar}).

The highest observed significance, with 8 observed events compared to 4.7 
expected background events 
($\xi$=0.95, $\sim$1.2$\sigma$), is at the location of the GeV blazar
3C273.
The second highest excess ($\xi$=0.84, $\sim$1.1$\sigma$) 
is from the direction of the Crab Nebula,
with 10 observed events compared to 6.7 
expected background events.
\begin{figure}[!t]
\begin{center}
\includegraphics[width=6.5in]{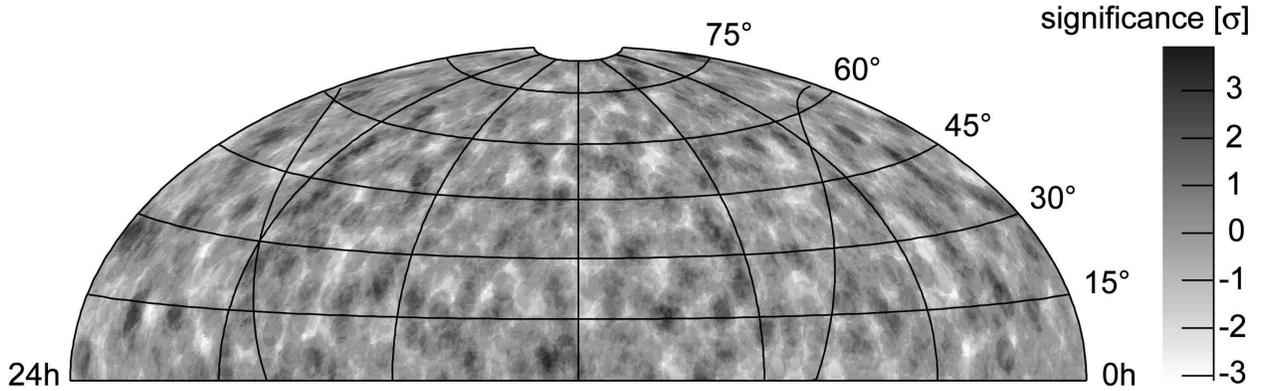}
\end{center}
\vskip -0.6cm
\caption{Sky map of the significance obtained by scanning of the northern
sky to search for event clusters. The significance is positive for 
excesses and negative for deficits of events compared to the expected 
background.}
\label{Fig:SkySig}
\end{figure}

\renewcommand{\arraystretch}{0.65}
\begin{table*}[htp]
\centering
  \begin{tabular}{le{2.1}e{2.1}e{1.2}ce{1.1}ce{2.1}e{1.2}e{1.2}e{2.1}e{2.1}e{1.2}e{1.2}e{3.0}}
  
&&&&&&& \multicolumn{4}{c}{\scriptsize $\begin{array}{cc} \gamma=2 \\ \overbrace{\hspace{4.0cm}} \end{array}$} 
     & \multicolumn{4}{c}{\scriptsize $\begin{array}{cc} \gamma=3 \\ \overbrace{\hspace{4.0cm}} \end{array}$} \\[-0.18cm]

Candidate    
&\talignc{$\delta$} 
&\talignc{$\alpha$}  
&\talignc{$r$} 
&\talignc{$n_{\mathrm{obs}}$} 
&\talignc{$n_\mathrm{b}$}     
&\hspace{0.08cm}
&\talignc{${\mu}_\mathrm{90}$}
&\talignc{${s}_{\nu_\mu}$}
&\talignc{${s}_{\nu_\tau}$}
&\talignc{$\mathrm{\Phi_{\nu_\mu}^0 + \Phi_{\nu_\tau}^0}$}
&\talignc{${\mu}_\mathrm{90}$}
&\talignc{${s}_{\nu_\mu}$}
&\talignc{${s}_{\nu_\tau}$}
&\talignc{$\mathrm{\Phi_{\nu_\mu}^0 + \Phi_{\nu_\tau}^0}$}
\\\hline
   \multicolumn{15}{c}{ \emph{TeV blazars} } \\\hline
Markarian 421 &38.2 &11.1 &3.25 &6 &7.4 &&4.1 &0.97 &0.15 &7.4 & 4.1 &0.15 &0.01 &51\\
Markarian 501 &39.8 &16.9 &3.00 &8 &6.4 &&7.9 &0.93 &0.14 &14.7& 8.3 &0.15 &0.01 &102 \\
1ES 1426+428  &42.7 &14.5 &2.75 &5 &5.5 &&4.8 &0.90 &0.13 & 9.4& 4.8 &0.16 &0.01 &58\\
1ES 2344+514  &51.7 &23.8 &2.50 &4 &6.2 &&3.1 &0.89 &0.15 & 5.9& 3.1 &0.19 &0.01 &29\\
1ES 1959+650  &65.1 &20.0 &2.25 &5 &4.8 &&5.6 &0.71 &0.11 &13.5& 5.6 &0.21 &0.02 &48 \\  \hline
   \multicolumn{15}{c}{ \emph{GeV blazars} } \\\hline
3C 273        & 2.1 &12.5 &3.75 &8 &4.7 &&9.6 &0.96 &0.10 &18.0& 9.8 &0.04 &\talignc{$\sim0$} &427\\ 
QSO 0528+134  &13.4 & 5.5 &3.50 &4 &6.1 &&3.2 &1.06 &0.14 & 5.3& 3.2 &0.08 &0.01 &72\\
QSO 0235+164  &16.6 & 2.6 &3.50 &7 &6.1 &&6.7 &1.03 &0.14 &11.4& 7.1 &0.09 &0.01 &145\\
QSO 1611+343  &34.4 &16.2 &3.25 &6 &7.0 &&4.5 &0.95 &0.15 & 8.3& 4.8 &0.14 &0.01 &65\\
QSO 1633+382  &38.2 &16.6 &3.25 &9 &7.4 &&8.1 &0.97 &0.15 &14.6& 8.3 &0.15 &0.01 &103 \\
QSO 0219+428  &42.9 & 2.4 &2.75 &5 &5.5 &&4.9 &0.89 &0.13 & 9.6& 4.8 &0.16 &0.01 &58\\
QSO 0954+556  &55.0 & 9.9 &2.50 &2 &6.7 &&1.4 &0.91 &0.15 & 2.7& 1.4 &0.20 &0.01 &12\\
QSO 0716+714  &71.3 & 7.4 &2.25 &1 &4.0 &&1.2 &0.70 &0.13 & 3.0& 1.2 &0.20 &0.02 &11\\ \hline
   \multicolumn{15}{c}{ \emph{Other AGNs} } \\\hline
M 87          &12.4 &12.5 &3.50 &6 &6.1 &&5.3 &1.07 &0.14 & 8.7& 5.7 &0.08 &0.01 &134\\
NGC 1275      &41.5 & 3.3 &3.00 &4 &6.8 &&2.7 &0.95 &0.14 & 5.0& 2.8 &0.16 &0.01 &31\\ \hline
   \multicolumn{15}{c}{ \emph{Micro-quasars and neutron star binaries} } \\\hline
SS433         & 5.0 &19.2 &3.75 &4 &6.1 &&3.1 &1.16 &0.13 & 4.8& 3.1 &0.06 &\talignc{$\sim0$} &96\\
GRS 1915+105  &10.9 &19.3 &3.50 &7 &6.1 &&6.8 &1.08 &0.14 &11.2& 7.1 &0.07 &\talignc{$\sim0$} &184\\
AO 0535+26    &26.3 & 5.7 &3.50 &7 &6.5 &&6.4 &0.99 &0.14 &11.3& 6.7 &0.11 &0.01 &112\\
GRO J0422+32  &32.9 & 4.4 &3.25 &9 &6.7 &&9.0 &0.94 &0.14 &16.7& 9.0 &0.14 &0.01 &123\\
Cygnus X-1    &35.2 &20.0 &3.25 &8 &7.0 &&7.3 &0.95 &0.15 &13.2& 7.3 &0.14 &0.01 &96\\
Cygnus X-3    &41.0 &20.5 &3.00 &7 &6.5 &&6.4 &0.95 &0.14 &11.8& 6.8 &0.16 &0.01 &80\\
XTE J1118+480 &48.0 &11.3 &2.75 &3 &7.1 &&1.5 &0.97 &0.14 & 2.8& 1.5 &0.19 &0.01 &15\\
CI Cam        &56.0 & 4.3 &2.50 &9 &6.3 &&9.4 &0.91 &0.14 &17.8& 9.5 &0.20 &0.01 &88\\
LS I +61 303  &61.2 & 2.7 &2.25 &5 &4.8 &&5.6 &0.75 &0.13 &12.6& 5.6 &0.20 &0.01 &50\\  \hline
   \multicolumn{15}{c}{ \emph{SNR and pulsars} }\\\hline
SGR 1900+14   & 9.3 &19.1 &3.50 &5 &5.7 &&4.8 &1.09 &0.13 & 7.8& 4.8 &0.07 &\talignc{$\sim0$} &127\\
Geminga       &17.9 & 6.6 &3.50 &3 &6.2 &&2.0 &1.01 &0.14 & 3.5& 2.0 &0.10 &0.01 &38\\ 
Crab Nebula   &22.0 & 5.6 &3.50 &10&6.7 &&10.1&0.98 &0.15 &17.8&10.4 &0.10 &0.01 &192\\
PSR 1951+32   &32.9 &19.9 &3.25 &4 &6.7 &&2.7 &0.94 &0.14 & 5.0& 2.7 &0.14 &0.01 &38 \\
Cassiopeia A  &58.8 &23.4 &2.50 &5 &6.0 &&4.4 &0.86 &0.13 & 8.9& 4.4 &0.20 &0.01 &41\\
PSR J0205+6449&64.8 & 2.1 &2.25 &1 &4.7 &&1.3 &0.72 &0.11 & 3.1& 1.3 &0.21 &0.02 &11\\ \hline
   \multicolumn{15}{c}{ \emph{Unidentified high energy gamma-ray sources} }\\\hline
3EG J0450+1105 &11.4 & 4.8 &3.50 &8 &5.9 &&8.4 &1.08 &0.14 &13.8& 8.6 &0.08 &\talignc{$\sim0$} &218 \\
TeV J2032+4131 &41.5 &20.5 &3.00 &7 &6.8 &&6.1 &0.95 &0.14 &11.2& 6.5 &0.16 &0.01 &76 \\ 
  \end{tabular}
  \caption{\label{Tab:32}
        Flux upper limits for selected neutrino source candidates:
        source directions
        (declination $\delta$ in degrees and right ascension 
        $\alpha$ in hours), search bin size ($r$ in degrees),
        number of observed events ($n_\mathrm{obs}$) and expected 
        background ($n_\mathrm{b}$). ${\mu}_\mathrm{90}$ is the event
        upper limit at 90\% CL (different for $\gamma$=2 and
	$\gamma$=3 because of different systematic errors) and ${s_{\nu_\mu}}$ (${s}_{\nu_\tau}$) is
	the expected number of events from muon (tau) neutrino and anti-neutrino
        interactions for a differential flux
	$\mathrm{\frac{d\Phi}{dE}=10^{-11}\cdot(\frac{E}{1\,TeV})}^{-\gamma}$
        $\mathrm{TeV}^{-1}\,\mathrm{cm}^{-2}\,\mathrm{s}^{-1}$.
        The 90\% CL upper limits 
        ($\mathrm{\Phi_{\nu_\mu}^0 + \Phi_{\nu_\tau}^0}$, neutrino and 
	anti-neutrinos) 
        are given in units of 
        $10^{-11}\,\mathrm{TeV}^{-1}\,\mathrm{cm}^{-2}\,\mathrm{s}^{-1}$, for both spectral indices $\gamma$=2 and $\gamma$=3.
        }
  \end{table*}
\renewcommand{\arraystretch}{1}

For the directions of all selected source candidates, 
the observations are compatible
with statistical fluctuations of the background. In Table~\ref{Tab:32}
we report flux upper limits for the spectral indices $\gamma$=2 and $\gamma$=3, 
following the Feldman and Cousins ordering
principle~\cite{feldman1998} and including a Bayesian treatment of 
systematic errors~\cite{hill2003b,conrad2003}.
The validity range of these limits -- here defined as the 
90\% energy containment 
region --
is between 1.6 TeV and 2.5 PeV for $\gamma$=2 and between 
0.1 TeV and 25 TeV 
for $\gamma$=3
(see Table~\ref{Tab:Ene}).  
The flux upper limits
on the sum of muon and tau neutrinos are calculated under the
assumption of a flavor ratio $\mathrm{\Phi_{\nu_\mu+\bar{\nu}_\mu}/\Phi_{\nu_\tau+\bar{\nu}_\tau}}$=1
at the earth. Please note that in previous publications of  
the AMANDA collaboration point source limits were presented only on the flux of muon neutrinos,
neglecting the sensitivity to tau neutrinos (i.e.~assuming $s_{\nu_\tau}=0$). Both limit representations, 
the former and the current one, can be calculated and converted into each other from the values of the 
event upper limit ${\mu}_\mathrm{90}$ and the expected number of signal events from muon~(tau) neutrinos 
and antineutrinos $s_{\nu_\mu}$~($s_{\nu_\tau}$) given in Table \ref{Tab:32}:

\begin{displaymath}
\mathrm{\Phi_{\nu_\mu+\bar{\nu}_\mu}^{0}}= \frac{1}{2} 
   \left( \mathrm{\Phi_{\nu_\mu+\bar{\nu}_\mu}^{0}+\Phi_{\nu_\tau+\bar{\nu}_\tau}^{0}}\right) =  
   \frac{{\mu}_\mathrm{90}}{s_{\nu_\mu}+s_{\nu_\tau}} 
   \, 10^{-11}\,\mathrm{TeV}^{-1}\,\mathrm{cm}^{-2}\,\mathrm{s}^{-1} \mathrm{,}
\end{displaymath} 

All limits include a systematic uncertainty of 15\% (20\%) in the signal
prediction for $\gamma$=2~(3) (see Section~\ref{Sec:Sys})
and a statistical error of 7\% in the background estimation.

\subsection{Northern sky survey}
\label{Sec:North}
A full scan of the northern sky is also performed
to look for any localized event
cluster, using a grid of circular search sky bins centered at distances of
$0.5^{\circ}$. The bin size is chosen according to the
optimization results reported in Section~\ref{Sec:Rec}. The strong 
bin correlation in this method ensures a high
detection chance without a high trial factor penalty and does not
require grid-shifts to account for boundary leaking
effects~\cite{hauschildt:phd}. 

The background per bin
is estimated in the same way as in the search for neutrinos
from known candidates, but a singularity arises at the pole, where
this method is not reliable. 
We therefore limited this search to
events with declination below $85^{\circ}$, giving 4251 remaining upgoing muons.
The statistical significance of any deviation 
is evaluated with the same technique
as above. 
Simulations with randomized events are performed using the same grid
of search bins, to account for  
the trial factor and the bin-to-bin correlations. 

Figure~\ref{Fig:SkySig} shows the distribution of observed
significances, in standard deviations. 
All 
observed excesses and deficits are compatible with statistical
fluctuations of background. The highest positive deviation corresponds to 
about 3.7$\sigma$. The probability of such a deviation or
higher due to background, estimated with 100 equivalent sky surveys of events with
randomized right ascension, is 69\%. 
\begin{figure}[!t]
\begin{center}
\includegraphics[width=6.5in]{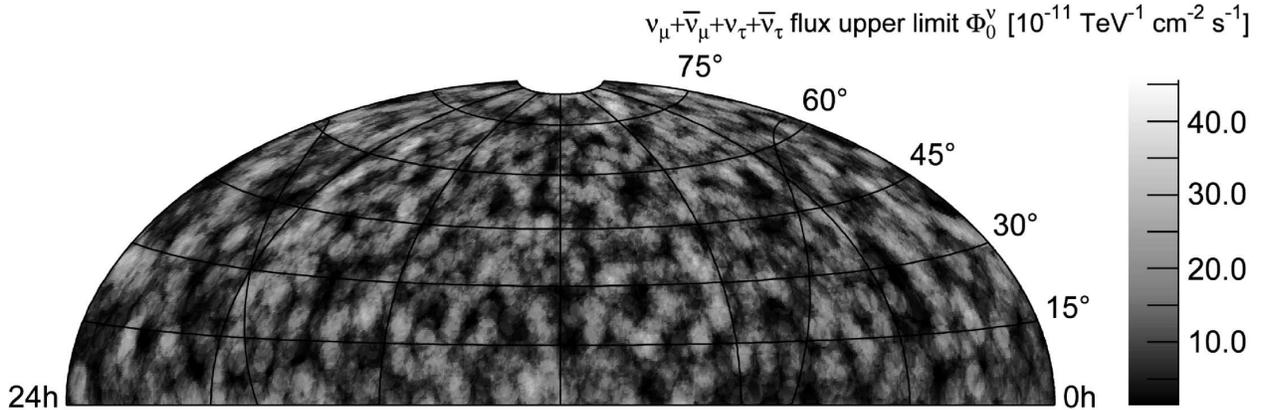}
\end{center}
\vspace{-0.5cm}
\caption{Map of the 90\% CL upper limits to the muon and tau neutrino flux from the survey of the northern
sky, for a spectral index $\gamma$=2. Limits are given on the
normalization factor $\Phi^\mathrm{0}$ to the flux
$\mathrm{\frac{d\Phi}{dE}={\Phi}^0\cdot(\frac{E}{1\,TeV})}^{-\gamma}$.}
\label{Fig:UpperL}
\end{figure}

Also in this case we give the neutrino flux upper limits as a function of
declination and right ascension including the systematic uncertainty
discussed in Section~\ref{Sec:Sys}. 
The results for a spectral index $\gamma$=2 are shown in
Fig.~\ref{Fig:UpperL}. The 
experimentally observed upper limit for $\gamma$=2 (averaged over declination and right
ascension) is 
$\mathrm{\Phi_{\nu_\mu+\bar{\nu}_\mu}^\mathrm{0}+\Phi_{\nu_\tau+\bar{\nu}_\tau}^\mathrm{0}}
= 10.6
\cdot 10^{-11} \,\mathrm{TeV}^{-1}\,\mathrm{cm}^{-2}\,\mathrm{s}^{-1}$,
without
systematic error and 
$\mathrm{\Phi_{\nu_\mu+\bar{\nu}_\mu}^\mathrm{0}+\Phi_{\nu_\tau+\bar{\nu}_\tau}^\mathrm{0}}
= 11.1
\cdot 10^{-11} \,\mathrm{TeV}^{-1}\,\mathrm{cm}^{-2}\,\mathrm{s}^{-1}$
with systematic error.  
The first can be compared to the sensitivity
($\mathrm{\Phi_{\nu_\mu+\bar{\nu}_\mu}^\mathrm{0}+\Phi_{\nu_\tau+\bar{\nu}_\tau}^\mathrm{0}}
= 10.0 \cdot 10^{-11} \,\mathrm{TeV}^{-1}\,\mathrm{cm}^{-2}\,\mathrm{s}^{-1}$),
which gives the
expected average upper limit before the experimental observations
are performed. The agreement between the two quantities is an
independent confirmation of the compatibility of the experimental
observations with the estimated background.

The results of this sky survey are compared to those from an
analysis using an independent cluster search algorithm that is 
based on an un-binned likelihood procedure in which the events
are weighted with the individual track angular 
resolution~\cite{neunhoeffer2006b}. The two methods
yield consistent results for the significance map.

\section{Search for spatially correlated and cumulative excesses}
\label{Sec:Stack-Corr}
As no statistically significant accumulation of events could be
established in the sky map, 
two searches for cumulative effects are also made. 
The first search aims at detecting 
correlations between spatial event coordinates, independent of source 
candidates; the second tests
the cumulative significance of pre-defined catalogs of 
objects (source stacking).
The purpose is to search for an excess 
of events from the sum of several sources, where the individual
fluxes are below the detector sensitivity but the integrated signal yields a significant
excess over the background.

\subsection{Search for an excess at small event separation angles}
The combined effect from several weak sources can be
observed as an excess of event pairs at a small angular
distance, consistent with the point spread function of the
detector. 
The number of individual sources and their location would remain
undefined.

We look for angular correlations in the final event sample
by comparing the distribution of the squared separation angle of 
event pairs to a template distribution
expected for atmospheric neutrinos. The latter is
obtained from 10000 neutrino samples randomizing the right
ascension of the 4282 observed events. This properly takes into account the
expected declination distribution, which is not uniform because 
of the effective area and the angular distribution of primary cosmic rays
\footnote{Please note that this randomization technique implies that the 
statistical test described below is insensitive to potential sources at declinations higher than $\delta \approx 85^{\circ}.$}.

\begin{figure}[!t]
\begin{center}
\includegraphics[width=4in]{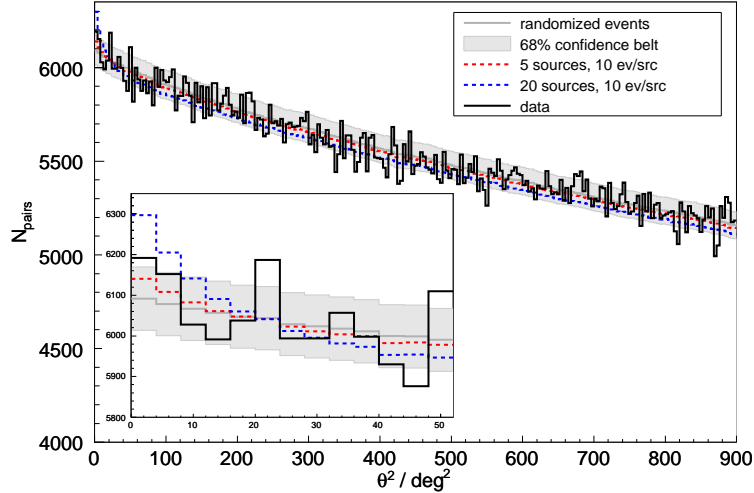}
\end{center}
\vskip -0.6cm
\caption{Distribution of the separation angle between event pairs for
the experimental data
and for two samples consisting of the original data plus signal events
following the point
spread function of the detector (5 and 20 sources respectively, 
each contributing
10 neutrinos on average). 
The distribution expected for a pure atmospheric
sample is also shown (randomized events) with a 1$\sigma$
confidence belt.}
\label{Fig:Corr}
\end{figure}

The distribution of the separation angle between event pairs is shown
in Figure~\ref{Fig:Corr}, with the corresponding 1$\sigma$ confidence belt. The
potential to identify a signal contribution with this method is also
shown for the cases of 5 and 20 sources respectively, each contributing on average 10 neutrinos to
the final event sample. An accumulation at small
separation angles would be expected.

In order to detect a signal at a 3$\sigma$ level, a reduced $\chi^2$ value
of more than 1.31 should be found in the comparison of the data 
with the template.
The reduced $\chi^2$ value obtained with the observed event sample 
is 203/223 for 223 degrees of
freedom and the probability to obtain a larger or equal value is 
81\%. No indications of a contribution from an ensemble of 
weak neutrino sources is therefore found by this test.
The probability of a 3$\sigma$ detection would be  
11\% and 65\% for 5 and 20 sources, respectively.

In an alternative approach the sky-map is tested for correlated events
and distortions of the angular distribution resulting from many weak
sources. The sky-map is decomposed by means of spherical
harmonics and the power spectrum of their coefficients is analyzed.
This study will be presented elsewhere.

\subsection{Source stacking results}
The final data set has been searched for a signal due to the cumulative neutrino
flux from generic classes of active galactic nuclei. 
Recently, a source stacking analysis was performed on data
collected in the year 2000~\cite{achterberg2006:stacking,ahrens2004:point}, defining
 samples of 10 classes of active galactic nuclei, according to
phenomenological selection criteria. 
The event sample extracted in this work yields
an increase in sensitivity by a factor $4$ 
compared to~\cite{achterberg2006:stacking,ahrens2004:point}.

The number of sources in each class was optimized assuming
a linear correlation between the photon and the neutrino flux.
The size of circular
search bins was optimized according to the point spread function of
this analysis. 

For each source class 
sample, the cumulative signal and the background expectation are
determined as the sum of the 
corresponding quantities of the individual sources. 
To ensure a statistically correct treatment of overlapping search
bins, events in the overlap area contribute only once to
the  cumulative signal. The background estimation is also corrected
for the bin correlations.

None of the classes tested show a statistically significant
excess over the background expectations.
Table~\ref{Tab:Stc} reports the 90\% confidence level upper limits 
to the cumulative neutrino flux, following the Feldman and Cousins ordering
principle~\cite{feldman1998} and including a Bayesian treatment of 
systematic errors~\cite{hill2003b,conrad2003}.
The limits include a systematic error of 15\% in the signal
prediction (see Section~\ref{Sec:Sys})
and a statistical error in the background estimation between 3\% and
7\%, depending on declination (Section~\ref{Sec:SteadyPoint}).
The limits do not include the contribution from tau neutrinos.
\renewcommand{\arraystretch}{0.7}
\begin{table*}[!h]
\centering
\begin{tabular}{r|ce{2.1}e{2.1}e{2.1}e{2.1}e{1.2}}
AGN class
&\talignc{$N_\mathrm{src}$} 
&\talignc{$n_\mathrm{obs}$} 
&\talignc{$n_\mathrm{b}$} 
&\talignc{${\mu}_\mathrm{90}$} 
&\talignc{$\Phi_{\nu_\mu}^\mathrm{0}$} 
&\talignc{$\Phi_{\nu_\mu}^\mathrm{0}/N_\mathrm{src}$} \\
\hline
GeV blazars              & 8  & 17 & 25.7 &2.7  &2.7 &0.34 \\
unidentified GeV sources & 22 & 75 & 77.5 &14.1 &16.5 &0.75\\
IR blazars               & 11 & 40 & 43.0 &9.3  &10.6 &0.96\\
keV blazars (HEAO-A)     & 3  & 9  & 14.0 &2.7  &3.6 &1.18\\
keV blazars (ROSAT)      & 8  & 31 & 33.4 &8.3  &9.7 &1.20\\
TeV blazars              & 5  & 19 & 23.6 &4.7  &5.5 &1.11\\
GPS and CSS              & 8  & 24 & 29.5 &5.0  &5.9 &0.74\\
FR-I galaxies            & 1  & 3  & 3.1  &4.3  &4.1 &4.11\\
FR-I without M87         & 17 & 40 & 57.2 &2.7  &2.9 &0.17\\
FR-II galaxies           & 17 & 77 & 68.5 &25.5  &30.4 &1.79\\
radio-weak quasars       & 11 & 35 & 41.6 &5.6  &6.7 &0.61\\
\end{tabular}
\caption{\label{Tab:Stc}Results of the stacking analysis for
each AGN class: 
number of sources ($N_\mathrm{src}$), number of
expected background events ($n_\mathrm{b}$) and number of observed
events ($n_\mathrm{obs}$).
${\mu}_\mathrm{90}$ is the cumulative event upper limit
and $\mathrm{\Phi_{\nu_\mu}^\mathrm{0}}$ is the upper limit to
the cumulative muon flux, in units of 
$10^{-11}\,\mathrm{TeV}^{-1}\,\mathrm{cm}^{-2}\,\mathrm{s}^{-1}$,
        for a spectral index $\gamma$=2.
The last column gives the
limits divided by the number of sources
($\mathrm{\Phi_{\nu_\mu}^\mathrm{0}}/N_\mathrm{src}$). These limits do
not include the contribution of tau neutrinos.}
\end{table*}
\renewcommand{\arraystretch}{1}

\section{Summary}
\label{Sec:Conc}
We have performed a search for point sources
of high energy neutrinos in the northern sky with the data collected with the 
AMANDA-II telescope in the years 2000 to 2004. Improved event
reconstruction and selection techniques have been applied,
with special emphasis on the
energy spectrum of the Monte Carlo events passing the selection cuts and
aiming at good sensitivity to a large variety of possible
signal energy spectra.

We selected the largest event sample ever extracted from data
collected with a neutrino telescope, 
consisting of 4282 upward going muon tracks
with good reconstruction quality. This is
in agreement with a Monte Carlo simulation of atmospheric neutrinos
yielding 4600$^{+300}_{-1000}$(sys) events
(Figure~\ref{Fig:Zenith}). The effects of oscillations of atmospheric neutrinos, yielding
muon neutrino disappearance and tau neutrino appearance are
negligible and the contribution of tau neutrinos to this event sample is below 0.5\%.

In contrast to that, we emphasize that in case of cosmic neutrinos a 
contribution from tau neutrinos is
expected in the sample of up-going muon tracks selected in this
analysis, depending on declination and energy spectrum.
We therefore
also estimated the contribution from charged current interactions of tau
neutrinos followed by tau decay into a muon (with a 17.7\% branching ratio).
Under the assumption
$\mathrm{\Phi_{\nu_\mu+\bar{\nu}_\mu}/\Phi_{\nu_\tau+\bar{\nu}_\tau}}$=1
at the earth, the additional contribution to the event sample from tau neutrinos, 
ranges from 10\% to 16\% for $\gamma$=2, depending on declination.

The sensitivity to a point source flux of muon and tau neutrinos (and anti-neutrinos)
is $\mathrm{\Phi_{\nu_\mu+\bar{\nu}_\mu}^\mathrm{0}+\Phi_{\nu_\tau+\bar{\nu}_\tau}^\mathrm{0}}
= 10.0 \cdot 10^{-11} \,\mathrm{TeV}^{-1}\,\mathrm{cm}^{-2}\,\mathrm{s}^{-1}$,
for 1001 effective days of exposure, in the energy range between 1.6 TeV and 2.5 PeV and assuming a flavor
ratio at earth of $\mathrm{\Phi_{\nu_\mu+\bar{\nu}_\mu}^\mathrm{0}/\Phi_{\nu_\tau+\bar{\nu}_\tau}^\mathrm{0}}=1$.
This is the declination-averaged sensitivity on the normalization factor 
$\Phi^\mathrm{0}$ to the flux
$\mathrm{\frac{d\Phi}{dE}={\Phi}^0\cdot(\frac{E}{1\,TeV})}^{-\gamma}$, assuming
$\gamma$=2. 
The representation is different from our previous papers,
where the sensitivity was given to the integrated muon neutrino flux and additionally 
the contribution from tau neutrinos was neglected. For this work the 
sensitivity to the muon neutrino component of the flux would be 
$\mathrm{\Phi_{\nu_\mu+\bar{\nu}_\mu}^\mathrm{0}
=1/2 \, (\Phi_{\nu_\mu+\bar{\nu}_\mu}^\mathrm{0}+\Phi_{\nu_\tau+\bar{\nu}_\tau}^\mathrm{0})}= 
5.0 \cdot 10^{-11} \,\mathrm{TeV}^{-1}\,\mathrm{cm}^{-2}\,\mathrm{s}^{-1}$.
 
The average experimentally observed upper limit of this analysis for $\gamma$=2,
averaged over declination and right ascension, is  
$\mathrm{\Phi_{\nu_\mu+\bar{\nu}_\mu}^\mathrm{0}+\Phi_{\nu_\tau+\bar{\nu}_\tau}^\mathrm{0}}
= 10.6
\cdot 10^{-11} \,\mathrm{TeV}^{-1}\,\mathrm{cm}^{-2}\,\mathrm{s}^{-1}$, without
systematic error and 
$\mathrm{\Phi_{\nu_\mu+\bar{\nu}_\mu}^\mathrm{0}+\Phi_{\nu_\tau+\bar{\nu}_\tau}^\mathrm{0}}
= 11.1
\cdot 10^{-11} \,\mathrm{TeV}^{-1}\,\mathrm{cm}^{-2}\,\mathrm{s}^{-1}$, including the
systematic error, consistent with the expected sensitivity.
An overall improvement of
approximately four (1.5) times is achieved compared to the sensitivity after 
197 (607) days of
exposure~\cite{ahrens2004:point,ackermann2005:point}, when neglecting
the sensitivity to tau neutrinos in this comparison.
The improvement is due to the longer
live-time and a refined
event selection yielding a higher signal
efficiency and a better background rejection power.

We searched the sample of 4282 up-going muon tracks for a signal of
cosmic origin, testing first individual
directions. 
Table~\ref{Tab:32} reports the results of the searches on a catalog
of 32 selected sources. The highest excess, with a pre-trial
significance of 1.2$\sigma$, corresponds to the direction of the 
blazar 3C 273. The highest excess from the full northern sky
corresponds to a pre-trial significance of 3.7$\sigma$ (Figure~\ref{Fig:SkySig}).
The probability of such a deviation or
higher due to background is 69\%.
We also performed a search based on the angular separation of the
events (Figure~\ref{Fig:Corr}) and a stacking analysis of selected
active galactic nuclei
(Table~\ref{Tab:Stc}). 

No indication
of point sources of neutrinos was found. Therefore we present flux upper limits for the 32 source candidates,
assuming spectral indices $\gamma$=2 and $\gamma$=3, at a 90\%
confidence level (Table~\ref{Tab:32}). Additionally we provide a map
of the flux upper limit for $\gamma$=2 for the full northern sky.
We give the most stringent flux upper limits to date.

\section{Acknowledgments}
 We acknowledge the support from the following agencies: National Science Foundation-Office of Polar Program, National Science Foundation-Physics Division, University of Wisconsin Alumni Research Foundation, Department of Energy, and National Energy Research Scientific Computing Center (supported by the Office of Energy Research of the Department of Energy), the NSF-supported TeraGrid system at the San Diego Supercomputer Center (SDSC), and the National Center for Supercomputing Applications (NCSA); Swedish Research Council, Swedish Polar Research Secretariat, and Knut and Alice Wallenberg Foundation, Sweden; German Ministry for Education and Research, Deutsche Forschungsgemeinschaft (DFG), Germany; Fund for Scientific Research (FNRS-FWO), Flanders Institute to encourage scientific and technological research in industry (IWT), Belgian Federal Office for Scientific, Technical and Cultural affairs (OSTC); the Netherlands Organisation for Scientific Research (NWO); M. Ribordy acknowledges the support of the SNF (Switzerland); J. D. Zornoza acknowledges the Marie Curie OIF Program (contract 007921). 

\bibliographystyle{apsrev}
\bibliography{PointSource5yrIntegrated_v1.9}

\end{document}